\documentclass[journal,letterpaper]{IEEEtran}

\usepackage{amsmath,amssymb,amsfonts}
\usepackage{amsthm}

\usepackage{algorithmic}
\usepackage{graphicx}
\usepackage{booktabs}
\usepackage{multirow}
\usepackage{cite}
\usepackage{xcolor}
\newcommand{\new}[1]{#1}
\newcommand{\rev}[1]{#1}
\usepackage{url}
\usepackage{textcomp}
\usepackage{hyperref}
\usepackage{enumitem}
\setlist[enumerate]{leftmargin=*, labelindent=0pt, itemsep=1pt, topsep=3pt,
 parsep=0pt, partopsep=0pt}
\setlist[itemize]{leftmargin=*, labelindent=0pt, itemsep=1pt, topsep=3pt,
 parsep=0pt, partopsep=0pt}
\setlist[description]{leftmargin=0pt, labelindent=0pt, itemsep=1pt, topsep=3pt,
 parsep=0pt, partopsep=0pt, style=unboxed,
 font=\normalfont\bfseries}

\graphicspath{{figures/}}

\begin{document}

\title{Scheduling of Variable-Speed Pumped Storage via Hierarchical Reinforcement Learning
\author{SangWoo~Park,~\IEEEmembership{Member,~IEEE,}
 Kyung-bin~Kwon,~\IEEEmembership{Member,~IEEE,}
 and~Dam~Kim,~\IEEEmembership{Member,~IEEE}%
\thanks{\emph{\rev{All authors contributed equally to this work.} (Corresponding author: SangWoo Park)} \textbf{S.~Park} is with the Department of Mechanical and Industrial
Engineering, New Jersey Institute of Technology, Newark, NJ 07103, USA
(e-mail: sangwoo.park@njit.edu). \textbf{K.~Kwon} is with the Optimization and Control
Group, Pacific Northwest National Laboratory, Richland, WA 99352, USA
(e-mail: kyungbin.kwon@pnnl.gov). \textbf{D.~Kim} is with the Department of Convergence
System Engineering, Chungnam National University, Daejeon 34134, South Korea
(e-mail: damkim@cnu.ac.kr).}%
}
}

\markboth{IEEE Transactions on Sustainable Energy, Submitted}{}

\maketitle

\begin{abstract}
Variable-speed pumped storage hydropower (VSPSH) plants modulate power
continuously in both generation and pumping, a match for the net-load ramps of
rising wind and solar. Their flexibility is bounded by hydraulics rather than
by a power rating, so the deliverable power range shifts with net head and
excludes small outputs entirely. This paper addresses bidding and dispatch for
such a plant across a two-settlement market with a \emph{hierarchical
reinforcement-learning} formulation. Its two layers are circularly dependent: a
real-time (RT) bid is defined only against a cleared day-ahead (DA) commitment,
and the DA reward only against an RT dispatch policy. Our proposed
\emph{feasibility-constrained domain randomization} breaks the cycle by
training the RT layer on generated deliverable DA schedules before any bidding
policy exists; the DA layer then optimizes against that frozen RT behavior, and
the two alternate. Learning a policy at once profitable and deliverable is
obstructed by the nonconvex operating region: penalizing infeasibility competes
against the revenue the offer earns, discouraging an undeliverable offer
without preventing one. We therefore pursue
safety through \emph{action-space modification} rather than through the reward,
combining action masking of the bid space, projection of each commitment before
settlement, and replacement by a fail-safe controller. \rev{Over 48
out-of-sample days it earns \$3{,}313/day more than an equivalent penalty
formulation while holding undeliverable commitment at a median of zero, which
the penalty formulation does not, and no ablation improves on it without giving
up deliverability or the storage target.}
\end{abstract}

\begin{IEEEkeywords}
\rev{Energy storage}, Pumped storage hydropower, variable-speed operation,
\new{hierarchical reinforcement learning.}
\end{IEEEkeywords}

\vspace{-4mm}
\section{Introduction}
\label{sec:intro}

Deep decarbonization is displacing dispatchable thermal generation with variable
wind and solar, and the resulting net-load ramps and price volatility place a
premium on bulk energy storage. Pumped storage hydropower (PSH) remains the
largest installed storage class worldwide, and in restructured systems its
economics are increasingly set by wholesale market arbitrage rather than by
regulated capacity payments \cite{koritarov2014modeling, perezdiaz2015trends}.
Conventional fixed-speed PSH serves that role poorly. Its synchronous machine is
mechanically locked to grid frequency, so generation is confined to a narrow band
around rated power and pumping is effectively an on/off decision. Variable-speed
pumped storage hydropower (VSPSH) removes the restriction by decoupling rotor
speed from grid frequency through a doubly-fed induction machine or a full
converter, permitting continuous modulation of power in both generation and
pump modes and improving part-load efficiency \cite{vasudevan2021variable,
bidgoli2020dfim, iliev2019variable, jukic2023optimal}. For a market
participant the essential gain is that pumping becomes a continuously biddable
quantity rather than a binary commitment.

That gain comes with a harder decision problem, because VSPSH flexibility is
bounded by hydraulics rather than by a power rating. The converter's slip range
bounds speed re-centering electrically, while cavitation, reverse-flow and
vibration limits bound it hydraulically, and both tighten away from the design
head \cite{vasudevan2021variable, iliev2019variable, bidgoli2020dfim}. The
feasible set is therefore \emph{head-dependent} and \emph{nonconvex}: a
generation branch, a pump branch, and an idle point, with both branches
contracting at the head extremes and excluding small outputs entirely. A
commitment deliverable at mid-reservoir may be unrealizable at the same MW level
once the head has moved, and the head is itself a consequence of earlier
dispatch, so feasibility is path-dependent.

Power-system scheduling has long recognized head-dependence. The canonical
mixed-integer linear programming (MILP)
treatment of short-term hydro unit commitment with a head-dependent reservoir
is \cite{borghetti2008milp}, extended in \cite{cheng2016hydro} to reservoirs
with multiple vibration zones, bands of output that must be permanently avoided
and whose position moves with net head. For PSH more broadly, existing
formulations schedule price-taking variable-speed plants for energy and
reserve, co-optimize mixed fixed- and variable-speed fleets under water-head
effects, embed variable-speed units in hydro-wind-solar portfolios, build risk-
and profit-based bidding strategies, and map head-dependent capability into
multi-segment ISO bid formats \cite{chazarra2018optimal, cao2023optimal,
jukic2023optimal, jukic2024optimal, wang2025short, tian2020risk, kwon2026stochastic}. These formulations are almost
exclusively day-ahead. The five-minute real-time market, where continuous
modulation is most valuable and where state of charge (SoC) actually evolves,
is represented as a balancing residual rather than as a second decision layer
with its own policy.

What a VSPSH plant may bid is set by tariff. Federal Energy Regulatory
Commission (FERC) Order No.~841 requires every U.S.\ regional transmission
organization or independent system operator (RTO/ISO) to adopt a storage
participation model in which a resource submits a discharge offer and a charge
bid, each as price--quantity pairs cleared against the locational marginal
price (LMP) \cite{ferc2018order841}. We adopt exactly this action space, so
bidding both price and quantity is the mechanism the tariffs provide rather
than a modeling convenience, though self-bidding is permitted rather than
universal \cite{pjm2023markets}. Settlement falls on the cleared quantity, and PJM does not
account for state of charge when dispatching a storage resource, leaving it to
the participant's own bidding parameters and offer prices \cite{pjm2026esr}.
The clearing carries no model of the head trajectory that the plant's own
earlier dispatch determines. A participant can protect itself by bidding only
what it could deliver at the worst head it might reach, but the turbine branch
contracts to $70\%$ of rated power there, so a bound safe at every head
forfeits much of what variable speed is installed to provide. Deliverability
therefore falls to the participant rather than to the mechanism.

Classical bidding methods formulate stochastic or robust programs over
price-scenario trees \cite{conejo2010decision, padhy2004unit}, and the
arbitrage value of storage policies has been studied
since \cite{sioshansi2009estimating}. The closest analogue
is \cite{krishnamurthy2018energy}, which finds a price--quantity bid earns
substantially more across the two settlements than a quantity-only bid. Its
asset is a battery with convex, SoC-independent power limits, so it does not
address what happens when a cleared quantity cannot be delivered.

Reinforcement learning (RL) offers a complementary route, since it learns a
policy against a simulator without requiring a tractable analytic model and
amortizes the online decision to a single forward pass. Existing RL work on
storage either learns real-time or day-ahead dispatch policies under price
uncertainty \cite{xu2019arbitrage, wang2018storage, zonjee2023energy}, couples
arbitrage to a battery degradation model \cite{cao2020deep}, learns genuine bid
objects such as high-dimensional monotone bid curves or temporally aware
multi-market bids \cite{liu2023high, li2024temporal}, or coordinates timescales
and risk preferences through multi-agent and hierarchical
designs \cite{ochoa2022multi, chen2026mars}. A hierarchy across the two
settlements is harder to train than the layer count suggests, because the
layers define each other: a real-time (RT) bid is meaningful only against a
cleared day-ahead (DA) commitment, and the DA reward is unknown until the RT
policy that will execute the commitment is fixed. Prior designs sidestep this
by letting a single planner set both timescales, or by training each layer
against a fixed heuristic for the other, and all of it assumes battery-like
feasibility. For hydro, RL has targeted reservoir operation rather than
bidding \cite{xu2021deep}, and for VSPSH the learning literature sits at the
device layer \cite{kwon2026degradation}. The VSPSH bidding literature is
optimization-based and day-ahead-centric, while the VSPSH learning literature
is control-theoretic and market-free. This paper occupies the gap between
them.

Placing a learned bidding policy in that gap exposes a problem the
battery-centric literature does not encounter. A battery's feasible set is a
convex interval, so clipping an infeasible request returns the nearest
feasible power; the VSPSH region contains an idle gap, so a sub-minimum
commitment is rewritten to zero and the plant is settled on a quantity it
never delivered. A penalty on that shortfall competes with the settlement it
is meant to discourage, so a learned policy can appear to satisfy the
constraint while systematically violating it. Deliverability is not the
only constraint the mechanism must carry. Storage has to reach its contracted
end-of-day level, which the day-ahead plan can only project and which the
real-time layer is free to undo as it trades. We therefore enforce
both through \emph{safety through action-space modification}, which
surveys of safe reinforcement learning for power systems and of provably safe
reinforcement learning both separate from penalty terms on exactly this ground
\cite{su2025review, krasowski2023provably}. This paper contributes the following.

\begin{enumerate}
\item A \emph{hierarchical bidding formulation} for VSPSH
(Sec.~\ref{sec:hrl_formulation}) that casts the two settlements as coupled
Markov decision processes (MDPs), an hourly DA bidding layer and a
five-minute RT dispatch layer, each
acting through the price--quantity bid the storage participation models
accept and each constrained by the head-dependent operating region.

\item \emph{Feasibility-constrained domain randomization}
(Sec.~\ref{sec:schedule_pipeline}), a constructive generator that produces
deliverable day-ahead schedules across seven behavior classes. It supplies the
RT layer's training distribution before any DA bidding layer exists, resolving
the circularity without joint optimization over both layers. The layers then
refine by \emph{anchored alternation}, each best-responding to the other
while the RT layer's training distribution grows by \emph{adding} its
partner's rollouts to the synthetic classes.

\item \emph{Safe scheduling and operation through action-space
modification} for both constraints,
instantiating three mechanisms, namely \emph{action masking} of the bid space onto the
operating region, \emph{action projection} of each commitment before
settlement (Sec.~\ref{sec:projection}), and \emph{action replacement} by a
fail-safe restoration controller (Sec.~\ref{sec:terminal_restore}).

\item A numerical study comparing the proposed method against alternative
configurations of each component
(Secs.~\ref{sec:ablations},~\ref{sec:rt_value}).
\end{enumerate}

Sec.~\ref{sec:background} reviews market mechanics and the VSPSH
operating region, Sec.~\ref{sec:methods} develops the plant model and the
feasibility mechanisms, Sec.~\ref{sec:results} reports the numerical study,
and Sec.~\ref{sec:conclusions} concludes. Every parameter value needed to reproduce the results, together with the per-class benchmark and the derivations omitted here for length, is given in the appendix of the preprint \cite{park2026vspsharxiv}.

\vspace{-3mm}
\section{Background}
\label{sec:background}

\vspace{-1mm}
\subsection{Two-Settlement Electricity Markets}

Most U.S. wholesale electricity markets (e.g., PJM, NYISO) operate a two-settlement structure consisting of a
day-ahead (DA) market and a real-time (RT) market
\cite{pjm2023markets,conejo2010decision}. This section fixes the
settlement notation used throughout.

The \emph{day-ahead market} clears once per day for the 24~hours of the
following operating day. Generators submit hourly supply offers
$(q_h^{\text{DA}}, p_h^{\text{DA}})$ for $h\!=\!0,\dots,23$ specifying a
quantity in MW and a marginal price in \$/MWh. Loads (and net buyers
such as pump-mode storage) submit corresponding demand bids. The
independent system operator (ISO) clears the market via a
security-constrained unit commitment, producing an hourly \emph{day-ahead locational marginal price (DA-LMP)},
$\pi_h^{\text{DA}}$. A bid is accepted when it is in-the-money relative to the clearing LMP; the cleared quantity is settled financially at the LMP the following day. Forward financial settlement occurs on the cleared DA quantity at $\pi_h^{\text{DA}}$, regardless of physical delivery.

The \emph{real-time market} clears every 5--15 minutes (5~min in PJM)
based on the security-constrained economic dispatch. Each interval $t$ produces a real-time locational marginal price (RT-LMP), $\pi_t^{\text{RT}}$. The market participant's
physical dispatch $P_t^{\text{disp}}$ may differ from its DA
commitment $C_h$ for the enclosing hour; the per-interval RT
settlement is the deviation $\delta_t = P_t^{\text{disp}} - C_h$
priced at $\pi_t^{\text{RT}}$. Execution with $\delta_t = 0$ at every
interval, in which the unit simply delivers its cleared schedule, is called
\emph{zero-deviation} and serves throughout as the reference against which the
real-time layer is measured. Total daily revenue is then

\vspace{-3mm}
\begin{equation}
\label{eq:two_settlement}
R = \underbrace{\sum_{h=0}^{23} C_h \pi_h^{\text{DA}}}_{\text{DA settlement}}
 + \underbrace{\sum_{t=0}^{T-1} \delta_t \pi_t^{\text{RT}} \Delta t}_{\text{RT settlement}}
 - R^{\text{op}}
\end{equation}

\vspace{-1mm}
\noindent where $\Delta t = 5/60$~h and $R^{\text{op}}$ is the
operating-region penalty of Sec.~\ref{sec:opclip_finding}. We model no
start-up cost and no ramping cost, so $R^{\text{op}}$ is the only term that
is not a market settlement, and under the mechanisms of
Sec.~\ref{sec:projection} it is zero throughout Sec.~\ref{sec:results};
reported revenue is therefore market cash.

By offering to generate above, or pump below, its cleared position, a
participant exploits RT-LMP movements away from the DA-LMP at the cost of
physical exposure. We call these \emph{upward} (UP) and \emph{downward} (DN)
deviation bids. How much to commit day-ahead and how much to leave to real time
is itself a decision, and for storage it is coupled across the day by the state
of charge.

\vspace{-3mm}
\subsection{Variable-Speed Pumped Storage Hydropower}
\label{sec:vsph_intro}
\vspace{-1mm}
A PSH plant uses a reversible turbine-pump unit
to cycle water between two reservoirs at different elevations. In
generation mode the upper reservoir feeds the turbine; in pump mode
electrical energy lifts water back to the upper reservoir~\cite{koritarov2014modeling}. The two modes convert between hydraulic power and electrical
power in opposite directions, and lose energy in both,

\vspace{-3mm}
\begin{equation}
\label{eq:hydropower}
P^{\text{gen}} = \eta(H, Q)\,\rho g |Q| H,
\qquad
P^{\text{pump}} = \frac{\rho g |Q| H}{\eta(H, Q)},
\end{equation}

\vspace{-1mm}
\noindent where $Q$ is the flow through the unit, $H$ the net hydraulic head between the
reservoirs, $\eta(H,Q) \le 1$ the unit's efficiency, and $\rho$ and $g$ the
fluid density and gravity. Generating delivers \emph{less} electrical power
than the water carries, while pumping consumes \emph{more} than it stores. The efficiency is jointly maximized at a single best
efficiency point $(H^*, Q^*)$; we approximate the surrounding hill
chart with a quadratic
\begin{equation}
\label{eq:efficiency}
\eta(H, Q) = \Pi_{\eta_{\min}}^{\eta_{\max}}\!\left(\eta_{\max} - c_H (H - H^*)^2 - c_Q (|Q| - Q^*)^2 \right)
\end{equation}

\vspace{-1mm}
\noindent where, here and throughout the paper,
\begin{equation}
\label{eq:proj}
\Pi_{a}^{b}(x) \triangleq \max\{a,\ \min\{b,\ x\}\}
\end{equation}
denotes the projection of a scalar onto the interval $[a,b]$. We write
$\Pi_{\mathcal{F}(H)}(\cdot)$ for the corresponding projection onto the
nonconvex operating region $\mathcal{F}(H)$ introduced below, which
Sec.~\ref{sec:projection} makes precise.

In addition to what was described in Sec.~\ref{sec:intro}, Eq.~\ref{eq:hydropower} adds a second reason for the head-dependent operating region. Because $P$ is linear in $H$ and the flow is bounded by the waterway and runner capacity, a given maximum flow yields less power as the reservoir drains. Peak generation therefore falls at low head for an energetic reason, separately from the slip and cavitation limits that shape the region elsewhere.

We characterize the boundary of this region by four piecewise-linear
functions of the normalized head fraction
$h_f = (H - H_{\min})/(H_{\max} - H_{\min}) \in [0, 1]$, namely the
turbine-branch limits $P_{\min}^{\text{turb}}(h_f)$,
$P_{\max}^{\text{turb}}(h_f)$ and the pump-branch limits
$P_{\min}^{\text{pump}}(h_f)$, $P_{\max}^{\text{pump}}(h_f)$.
The feasible operating region is then
\begin{equation}
\label{eq:opregion}
\mathcal{F}(H) = \{0\} \cup [P_{\min}^{\text{turb}}, P_{\max}^{\text{turb}}]
 \cup [-P_{\max}^{\text{pump}}, -P_{\min}^{\text{pump}}].
\end{equation}
None of the mechanisms developed in this paper depends on the particular
breakpoints. They require only the qualitative structure described next,
so the numerical values used in our experiments, which describe a
$100$\,MW-class unit, are deferred to the preprint.

\begin{figure}[t]
\centering
\includegraphics[width=0.8\linewidth]{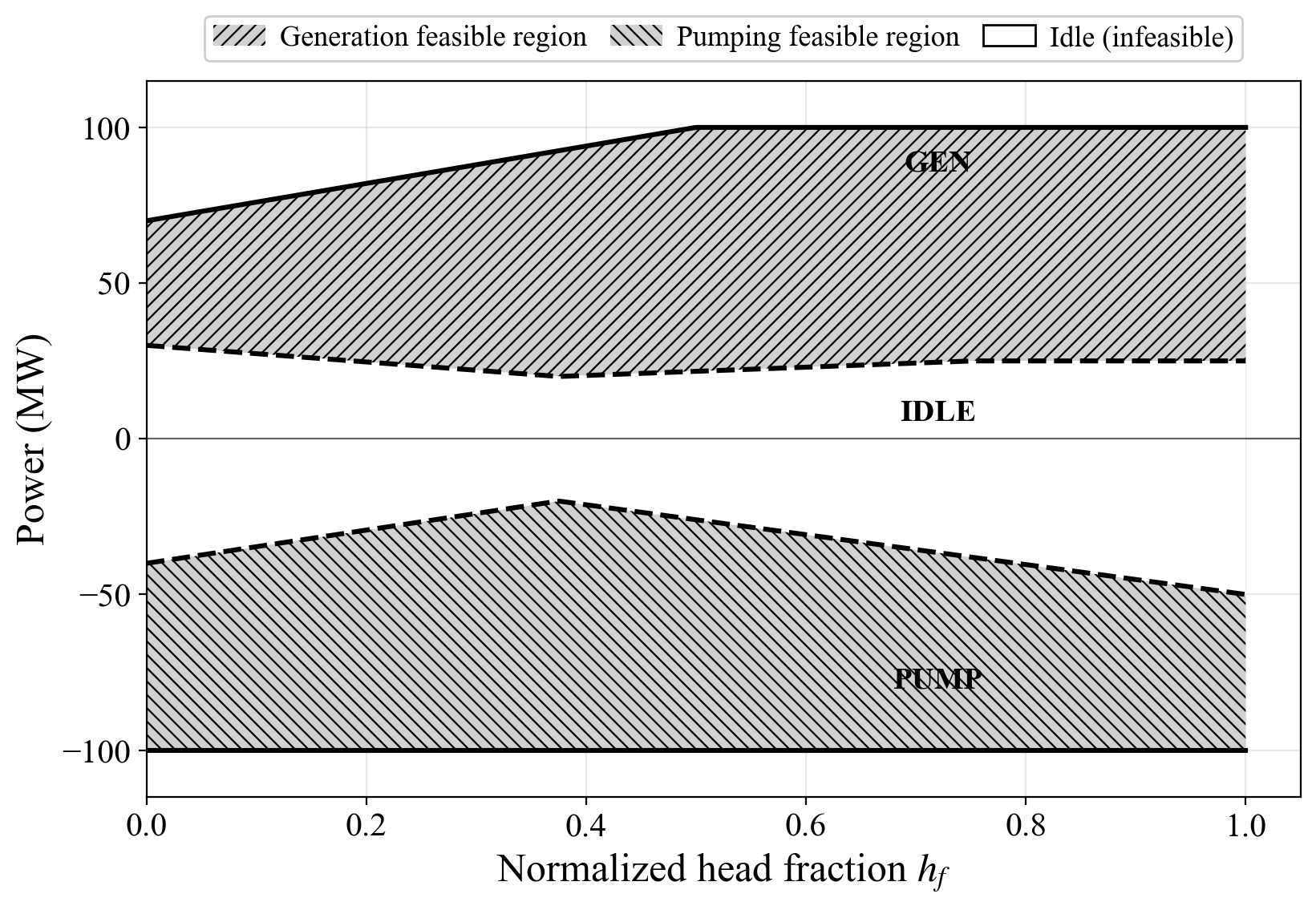}
\vspace{-3mm}
\caption{The feasible operating region $\mathcal{F}(H)$ of Eq.~\ref{eq:opregion}. Solid curves are $P_{\max}$, dashed curves $P_{\min}$.}
\label{fig:opregion}
\end{figure}

Figure~\ref{fig:opregion} renders the region. Three features matter for
bidding. First, the region is \emph{nonconvex}: the unit cannot sustain
small power magnitudes in either direction, so any request whose magnitude
falls below the relevant branch minimum can only be met by
idling. Second, both branches contract at the head extremes, so a
commitment that is feasible at moderate storage, with head near $H^*$, may
be infeasible at the same MW level when storage is nearly full or nearly
empty. Third, the attainable turbine maximum at low storage falls well
below the nameplate rating, so aggressive day-end commitments near
nameplate are physically unrealizable on days that drain the reservoir.

\vspace{-3mm}
\subsection{Why VSPSH Bidding is Hard}

Three structural features distinguish VSPSH bidding from conventional storage arbitrage:

\noindent \textbf{(i) Head-dependent feasibility:} Battery models abstract
feasibility into an SoC band and a constant power limit. A VSPSH plant has
both, plus the nonconvex operating region of Fig.~\ref{fig:opregion}, and a
commitment outside it cannot be followed in real time
(Sec.~\ref{sec:opclip_finding}).

\noindent \textbf{(ii) Hierarchical price exposure:} The DA agent commits an
hourly schedule under a DA-LMP forecast; the RT agent then dispatches at 5-min
resolution under the cleared $\pi_h^{\text{DA}}$ and a noisy
$\pi_t^{\text{RT}}$ forecast. The layers share one storage state and cannot be
optimized independently.

\noindent \textbf{(iii) Storage as a coupling variable:} Each commitment
changes the achievable head trajectory, and with it the operating region of
every later hour. The DA agent must reason about that trajectory; the RT agent
must exploit price deviations within the one it inherits.

\vspace{-1mm}
\section{Methods}
\label{sec:methods}

\vspace{-1mm}
\subsection{Plant Physical Model}
\label{sec:plant_model}

The unit we model is rated $100$\,MW in both generation and pumping.
Its usable reservoir spans $1\!\times\!10^{6}$ to $3\!\times\!10^{6}$\,m$^3$
over a $50{,}000$\,m$^2$ surface, so the geometric head is affine in storage
and sweeps $[80, 120]$\,m across the usable band. That is roughly
$490$\,MWh of usable energy, or a $4.9$-hour plant at rated power, and the
$50\%$ head swing is what makes the operating region move with the state of
charge. We model the plant at a 5-minute resolution. The reservoir storage
volume $V_t$ evolves under
\begin{equation}
\label{eq:storage}
V_{t+1} = \Pi_{V_{\min}}^{V_{\max}}\!\left(V_t + (Q^{\text{in}}_t - Q_t)\,\Delta t\right)
\end{equation}
where $Q_t$ is the flow through the unit (positive = generation,
negative = pumping), $Q^{\text{in}}_t$ is the natural inflow, and
$V_{\min}, V_{\max}$ are usable-storage bounds. We model the inflow as stochastic,
redrawn each interval about a constant mean; the agents' forecasts use that
mean, while the plant realizes the draw. The head is recovered geometrically
from $V_t$ and projected onto $[H_{\min}, H_{\max}]$.
Throughout the rest of the paper we work with the normalized state of
charge

\vspace{-2mm}
\begin{equation}
\label{eq:soc}
S_t = \frac{V_t - V_{\min}}{V_{\max} - V_{\min}} \in [0,1],
\end{equation}

\vspace{-1mm}
\noindent so that every storage quantity quoted later, including the end-of-day target
$S^{\text{tgt}}$ and the safety band, is a fraction of usable storage.
Power is additionally rate-limited: the dispatch may move by at most
$50$\,MW per five-minute step, \new{so the unit moves from idle to either
rating in two steps and reverses from rated pumping to rated generation in
four.}

Given a commanded power, Eq.~\ref{eq:hydropower} is solved for the flow $Q_t$
by fixed-point iteration, since $\eta$ itself depends on $Q$.
The hill chart is centred at $H^\star\!=\!100$\,m and
$Q^\star\!=\!113$\,m$^3$/s with $\eta_{\max}\!=\!0.90$, so over the
reachable operating region $\eta(H,Q)$ ranges from $0.79$ to $0.90$ with a
mean of $0.85$. Because $\eta$ multiplies the generating power and divides
the pumping power, the round-trip efficiency $\eta_{\text{pump}}\,
\eta_{\text{gen}}$ \new{runs from $0.72$ to $0.81$ at rated power depending on
head, and falls to $0.63$ at part load.} The instantaneous power is constrained to
$\mathcal{F}(H_t)$. An infeasible request is rewritten:
$P_t \notin \mathcal{F}(H_t)$ becomes either $0$ (when $|P_t|$ is below
the minimum) or $P_{\max}(H_t)$ in the matching branch (when $|P_t|$
exceeds the maximum). This rewriting matches the behavior of a real
plant operator who cannot dispatch at a setpoint outside the unit's
operating region, and is the source of the cleared-vs-executed gap we
discuss in Sec.~\ref{sec:opclip_finding}. A hard rail keeps storage inside $[0.05, 0.95]$ of the usable range. When
$S_t$ already sits below the lower rail the unit may not generate, and above
the upper rail it may not pump; the offending request is replaced by idle
rather than reduced in magnitude. No penalty term is attached to the
replacement, so the only cost of reaching a rail is the settlement the idled
intervals do not earn. The agent is warned before that happens by the shaping
term of Eq.~\ref{eq:safety_shape}, which charges a quadratic penalty on the
tighter band $[0.10, 0.90]$.

\vspace{-2mm}
\subsection{Hierarchical Bidding Formulation}
\label{sec:hrl_formulation}

\begin{figure}[t]
\centering
\includegraphics[width=0.75\linewidth]{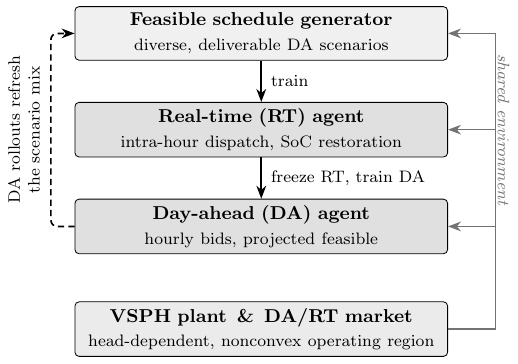}
\caption{Sequential co-training framework. \emph{Feasibility-constrained
domain randomization} supplies diverse, deliverable DA scenarios to the
real-time (RT) agent; the RT policy is then frozen and the day-ahead (DA)
agent trains against it. \emph{Anchored alternation} then refreshes the
RT training distribution by adding DA rollouts to the synthetic classes. }
\label{fig:framework}
\end{figure}

We decompose the joint bidding problem into a hierarchy of two coupled
MDPs: a \emph{day-ahead bidding layer} that
submits hourly price--quantity offers, and a \emph{real-time dispatch layer}
that operates the unit at five-minute resolution against the cleared schedule.
We refer to the trained policies as the DA and RT agents when discussing
training, and as the bidding and dispatch layers when discussing operation. Figure~\ref{fig:framework} gives an overview of the resulting training framework.

\smallskip
\noindent \textbf{(i) The dependency cycle:}
The RT agent's action depends on the DA commitment for the current
hour; the DA agent's reward depends on the RT agent's execution policy.
Write $C=(C_0,\dots,C_{23})$ for the day's vector of hourly DA
commitments, \new{$\varphi^{\text{DA}}$ for the day-ahead bidding policy that
produces them, and $\varphi^{\text{RT}}$ for the real-time execution policy.} Joint
optimization would require either an alternating fixed-point iteration or a
single agent acting over the \new{joint policy space
$(\varphi^{\text{DA}}, \varphi^{\text{RT}})$}, which is prohibitively high-dimensional. We instead adopt a sequential training recipe: first train the RT agent against a wide programmatic distribution of plausible DA commitments, then train the DA agent against the frozen RT policy.

\smallskip
\noindent \textbf{(ii) Robustness first, then synergy:}
\label{sec:why_iterate}
At round one no learned DA agent exists, and an RT action is defined only
relative to a commitment, so the RT agent must train against some
distribution of schedules. Any single DA policy would shape its dispatch
reflexes around that policy's idiosyncrasies, so we use a deliberately
broad, DA-agnostic distribution of feasible schedules
(Sec.~\ref{sec:schedule_pipeline}). The result is an executor that reacts
well to whatever schedule it is handed, and a frozen RT that is a stable
environment for the DA agent to explore against. Once a DA agent exists, its
own rollouts become the most relevant distribution the RT will face, so we
iterate: round two retrains the RT agent on the original seven classes plus
an eighth drawn from DA rollouts, then retrains the DA agent against the
updated RT. Round one buys \emph{robustness}, round two \emph{synergy}; Sec.~\ref{sec:ablations} measures the alternation and finds the round-2
gain positive.

We keep the seven synthetic classes in round two, against the intuition
that the dispatch layer will only ever execute schedules from its own bidding
layer. Training a round-2 dispatch policy exclusively on day-ahead rollouts
refutes that intuition: the specialized policy is worse overall and, decisively,
no better on day-ahead rollouts themselves
(Sec.~\ref{sec:results_rt_diversity}). The
synthetic classes act as a regularizer rather than as insurance bought at the
target distribution's expense.

\subsubsection{Real-time dispatch MDP}
The RT agent operates the unit over $T = 288$ five-minute steps. At
step $t$ the state $s_t \in \mathbb{R}^{72}$ collects head, storage,
the past four power dispatches, the next-12-step RT-LMP forecast, the
remaining hours of DA commitments and DA-LMPs, a storage-deficit
feature, \new{four clock features giving the hours remaining, hour of day, minute
within the hour and day of week, and the average energy cost of the storage
currently held.}
The agents reported in Sec.~\ref{sec:results} act on a reduced view of
this vector, with all but eight components zeroed. Head, storage, the near
real-time forecast, the current DA-LMP, the storage deficit and the hours
remaining survive, while the far forecast horizons and the later DA
commitments are suppressed. We chose the reduced view after comparing four
otherwise identical RT agents against a zero-deviation executor on the same
price days.
The action
$a_t = (\tilde q^{\text{UP}}_t, \tilde p^{\text{UP}}_t,
\tilde q^{\text{DN}}_t, \tilde p^{\text{DN}}_t) \in [-1, 1]^4$ encodes (quantity, price) pairs for an upward deviation offer and a
downward deviation bid. Quantities are decoded by linear maps; prices are decoded
as signed offsets from the hour's day-ahead price, which is the natural
reference point for a deviation bid,

\vspace{-5mm}
\begin{equation}
\label{eq:rt_bid_price}
p^{\text{UP}}_t = \pi^{\text{DA}}_h + \Delta_p^{\text{RT}}\,|\tilde p^{\text{UP}}_t|,
\qquad
p^{\text{DN}}_t = \pi^{\text{DA}}_h - \Delta_p^{\text{RT}}\,|\tilde p^{\text{DN}}_t|,
\end{equation}

\vspace{-1mm}
\noindent with $\Delta_p^{\text{RT}} = \$20$/MWh. The absolute values are not cosmetic. An
upward offer clears when $\pi^{\text{RT}}_t \ge p^{\text{UP}}_t$ and a downward bid
clears when $\pi^{\text{RT}}_t \le p^{\text{DN}}_t$, so the two are
well posed only when $p^{\text{UP}}_t \ge p^{\text{DN}}_t$. \rev{Under the unconstrained parameterization the opposite ordering is reachable, and there both orders clear on the same interval. Settlement is on the net deviation at the real-time LMP, so the two legs cancel and the interval produces no dispatch and no cash.} Half of the raw price plane therefore encodes pairs of orders whose net effect is nil, and within it many distinct actions map to the same outcome, so the settlement reward cannot distinguish them. An executor
trained without the constraint confirms this. Its median quotes invert to
$p^{\text{UP}} - p^{\text{DN}} = -\$13.16$/MWh, both orders clear
simultaneously on $28\%$ of discretionary intervals, and it sells into only
$26\%$ of the intervals where the real-time price exceeds the day-ahead price
by more than $\$3$/MWh, against $36\%$ for a fixed sensible quote. Forcing
$p^{\text{UP}}_t \ge \pi^{\text{DA}}_h \ge p^{\text{DN}}_t$ makes the
inverted half unreachable rather than merely unattractive.
The pair then clears against $\pi_t^{\text{RT}}$ to produce a deviation $\delta_t$ that
combines with the DA commitment $C_h$ to set the dispatch
$P_t^{\text{disp}} = \Pi_{C_h - \Delta_{\max}}^{\,C_h + \Delta_{\max}}(C_h + \delta_t)$, with
$\Delta_{\max} = 10$\,MW. The cap is a modeling choice rather than a
tariff rule. Real time settles the full deviation between metered output and
the cleared day-ahead position, and no ISO bounds its magnitude directly, so
$\Delta_{\max}$ stands in for the self-imposed limit a risk-managed
participant would place on its physical exposure.

The per-step reward is the unshaped settlement cash flow

\vspace{-2mm}
\begin{equation}
\label{eq:rt_reward}
r_t^{\text{RT}} = \delta_t \pi_t^{\text{RT}} \Delta t - \mathcal{P}^{\text{op}}_t
\end{equation}

\vspace{-1mm}
\noindent where $\mathcal{P}^{\text{op}}_t$ is a quadratic operating-region
penalty, zero inside $\mathcal{F}(H)$. The penalty is evaluated on the
requested power, before the operating region is applied, so it is a backstop
rather than the enforcement mechanism. In the proposed system the action mask
of Sec.~\ref{sec:projection} makes every request feasible and the term is
identically zero; it becomes active only in the ablations that remove the
mask, where it is the sole remaining defence.
A storage-shaping term may optionally be added during training. We use
a \emph{safety-only} shaping
\begin{equation}
\label{eq:safety_shape}
r_t^{\text{shape}} = -\mu_{\text{safety}}\!\left[\max(0, S^{\text{lo}} - \tilde S_t)^2 + \max(0, \tilde S_t - S^{\text{hi}})^2\right]
\end{equation}
with $S^{\text{lo}}\!=\!0.10$, $S^{\text{hi}}\!=\!0.90$,
$\mu_{\text{safety}}\!=\!50{,}000$. The penalty is zero everywhere
inside $[S^{\text{lo}}, S^{\text{hi}}]$; inside the safe range the
gradient comes purely from settlement. This contrasts with prior
trajectory-shaping designs that anchor the SoC to a per-episode
reference path; trajectory shaping is appropriate when training on a
single schedule shape but misfires on a diverse mix because the same
``deviate upward when above target'' reflex contradicts itself across schedule
types.

\smallskip
\subsubsection{Day-ahead bidding MDP}
\label{sec:da_mdp}
The DA agent operates a 24-step sequential MDP. At step $h$ it commits
one hour of dispatch via an action $a_h = (q_h, p_h) \in [-1, 1]^2$
encoding a (quantity, price) bid. The bid clears against
$\pi_h^{\text{DA}}$ via a uniform-pricing rule
\cite{conejo2010decision}: an in-the-money bid produces
$C_h = q_h$, an out-of-the-money bid produces $C_h = 0$. The price
component is parameterized as a delta from the DA forecast,
$p_h = \hat\pi_h^{\text{DA}} + (\text{tanh output})\cdot\Delta_p^{\text{DA}}$, so the
neutral action ($p_h = \hat\pi_h^{\text{DA}}$) clears with 50\%
probability and the agent need not learn the absolute price scale from
scratch.

The DA state $s_h \in \mathbb{R}^{75}$ collects the projected SoC, a one-hot encoding of the current hour, the
day's 24 DA-LMP forecasts, \rev{the commitments cleared in hours already
decided}, and the fraction of the day remaining,
with a day-of-week indicator. \rev{The sequential decomposition is a training
device: a participant submits all 24 offers before any of them clears, so a
deployable policy must form them from forecast information alone. Every number we
report is measured that way, with each hour's commitment assumed cleared
against the forecast rather than the realized price.} The DA per-hour reward is

\vspace{-3mm}
\begin{equation}
\label{eq:da_reward}
r_h^{\text{DA}} = C_h \pi_h^{\text{DA}}
 - \lambda_{\text{safe}} d_h^{\text{safe}}
 - \lambda_{\text{op}} d_h^{\text{op}}
\end{equation}

\vspace{-1mm}
\noindent plus, at $h\!=\!23$, the total RT settlement
$\sum_{t} \delta_t \pi_t^{\text{RT}} \Delta t$ and a terminal-SoC penalty.
The terminal penalty uses three coefficients. $\lambda_{\text{end}}$
and $\lambda_{\text{below}}$ weight the squared terminal error above and below
target, so the formulation can charge a shortfall more heavily than an
overshoot, and $\lambda_{\text{hinge}}$ weights a linear hinge applied in both
directions, so that any terminal error carries a non-zero gradient. Let $S^{\text{tgt}}$ denote the required end-of-day state of charge (we use
$S^{\text{tgt}}=0.25$) and $\tilde S_T$ the terminal SoC the DA plan projects
under zero-deviation execution. Writing $e = \tilde S_T - S^{\text{tgt}}$, the penalty is

\vspace{-5mm}
\begin{equation}
\label{eq:terminal_pen}
\lambda(e)\,e^2 + \lambda_{\text{hinge}}|e|,
\qquad
\lambda(e) = \begin{cases}
\lambda_{\text{below}}, & e < 0\\
\lambda_{\text{end}}, & e \ge 0,
\end{cases}
\end{equation}

\vspace{-1mm}
\noindent a quadratic well with a linear hinge. We set
$\lambda_{\text{below}} = \lambda_{\text{end}}$ throughout, so the well
reported here is symmetric and the hinge alone supplies the gradient at
target. \new{The penalty is paired with \emph{terminal-hour balancing}, which
overrides the hour-23 commitment with whatever magnitude lands $\tilde S_T =
S^{\text{tgt}}$ exactly under zero-deviation physics. Balancing carries the
constraint whenever that correction is itself deliverable, and the penalty
shapes hours 0--22 so that it usually is.} Here $d_h^{\text{safe}}$ is the MW
distance the storage-safety filter removes from the requested commitment,
and $d_h^{\text{op}}$ is the distance the operating-region filter rewrites,
\rev{summed over the twelve five-minute substeps of the hour}; both are
accumulated by simulating one hour of
zero-deviation dispatch under the same physics the RT environment
executes (Sec.~\ref{sec:opclip_finding}). The two feasibility penalties
are training-only signals, and backstops in the same sense, since the
projection of Sec.~\ref{sec:projection} drives $d_h^{\text{op}}$ to
approximately zero before they act; the eval metric reports the unshaped
settlement (Eq.~\ref{eq:two_settlement}).

A 20\%-probability \emph{adversarial DA-LMP} perturbation is applied
during training: when triggered, the realized $\pi_h^{\text{DA}}$ is
sampled from the convex hull of the forecast and the agent's bid price,
producing a perturbed clearing that punishes lazy bids placed far from
the forecast. This is not a market-design assumption but a training
device, designed to make the policy robust to the DA-LMP versus
forecast gap.

Both agents are trained with Soft Actor-Critic
\cite{haarnoja2018soft}; network sizes and optimizer settings are listed
in the preprint. To make best-checkpoint
selection and final performance reproducible rather than seed-dependent,
we normalize rewards online and clip gradients, and we evaluate on a
fixed held-out seed band, detailed in Sec.~\ref{sec:results_setup}.

\vspace{-2mm}
\subsection{Generating Deliverable Day-Ahead Schedules}
\label{sec:schedule_pipeline}
\vspace{-1mm}
The RT agent must train against a distribution of plausible DA
commitments. We construct this distribution programmatically from
eight schedule classes, each a stochastic generator parameterized by the LMP
forecast and the operating-region constants.
Four react to price: price-follow activates the most and least expensive
forecast hours, balanced pumps in the cheapest and generates in the most
expensive, peak-shave uses symmetric blocks in each direction sized by the
round-trip energy balance, and mixed-arbitrage takes that structure with
independent magnitudes. Three sample shape without reference to price, namely
random, feasible-arbitrage, and a flat all-idle day. The eighth draws from a
cache of $2{,}000$ schedules produced by a previously trained DA agent, built
with the operating-region projection active so that the cached schedules are
exactly the deliverable commitments the RT agent will face.

Each schedule is constructed to be deliverable rather than drawn and tested
afterwards. Given a class's intended shape, we forward-simulate storage hour by
hour and commit, at each hour, the largest magnitude in the intended direction
whose end-of-hour SoC remains inside $[0.07, 0.93]$. Because a constant-sign
commitment moves SoC monotonically within the hour, bounding the end-of-hour
value bounds the whole hour. The last two hours are set to whatever balances
SoC back to $S^{\text{tgt}}$, likewise projected onto the operating region.
This yields a deliverable schedule on essentially every draw; the flat class
alone remains all-idle, a deliberate no-position day.

A final step then overrides hour 23 of every accepted schedule with the magnitude that lands SoC exactly at $S^{\text{tgt}} = 0.25$ under zero-deviation physics, so that every training schedule ends at the same storage state and the agent cannot learn that an arbitrary ending is acceptable.

\vspace{-2mm}
\subsection{Deliverability of Cleared Commitments}
\label{sec:opclip_finding}

The DA and RT layers are settled on different quantities. The market settles
the commitment $C_h$ the DA agent submits, but the plant can only execute a
point of $\mathcal{F}(H_t)$, and the two need not coincide. A sub-minimum
commitment ($|C_h| < P_{\min}(H_h)$) clears at $\pi_h^{\text{DA}}$ and produces
a DA settlement, yet is executed as $0$~MW; the resulting
$|\delta_t| = |C_h|$ is repurchased at $\pi_t^{\text{RT}}$. Since
$\mathbb{E}[\pi_t^{\text{RT}}] \approx \pi_h^{\text{DA}}$, the net cash effect
is small and the commitment can even be profitable, so nothing in the
settlement itself discourages it.

This follows from the operating region meeting a settlement rule that
pays on the submitted quantity. \new{The gap can be priced, or it can be
prevented. Pricing it} means adding an operating-region rewrite-distance penalty

\vspace{-3mm}
\begin{equation}
\label{eq:opclip_pen}
\lambda_{\text{op}} d_h^{\text{op}}
\;\;\;\;\text{where}\;\;
d_h^{\text{op}} = \sum_{t \in h}\,\bigl|C_h - P_t^{\text{disp}}\bigr|
\end{equation}

\vspace{-1mm}
\noindent to the DA per-hour reward (Eq.~\ref{eq:da_reward}). Eliminating it means
projecting $C_h$ onto the operating region before settlement
(Sec.~\ref{sec:projection}). Eq.~\ref{eq:opclip_pen} places the responsibility on the learned
weights; we place it on the mechanism instead, and
Sec.~\ref{sec:penalty_to_projection} measures the difference.

\vspace{-2mm}
\subsection{Operating-Region Projection}
\label{sec:projection}

We project every DA commitment onto the operating region
before it is settled. Let $C_h$ be the storage-feasible commitment
requested at hour $h$ and $H_h$ the head at the hour's start. The
projection $\Pi_{\mathcal{F}(H_h)}(C_h)$ first maps the magnitude onto the
correct branch of $\mathcal{F}(H_h)$, so that a request below the branch
minimum is sent to $0$ and a request above the branch maximum is sent to
$P_{\max}(H_h)$. It then, because the head drifts within the hour as storage moves,
shrinks the magnitude by bisection until a one-hour zero-deviation
simulation rewrites no more than a small tolerance $\tau_{\text{op}}$
(we use $5$ MW-substeps/hour) of the commitment. The settled commitment
is $\Pi_{\mathcal{F}(H_h)}(C_h)$, so the DA agent is paid only for power it
can physically deliver, and the distance $d_h^{\text{op}}$ of Eq.~\ref{eq:opclip_pen}
is driven to approximately zero for every seed by the mechanism rather than
by the learned weights.
Two details preserve the rest of the design. First, the projection is
applied \emph{after} the storage-safety projection so it can only reduce
magnitude and never pushes storage toward a rail.
The projection has a companion that acts one step earlier. Rather than
letting the agent name any quantity and repairing it afterwards, the
\emph{action mask} maps the quantity component of the action onto the feasible
set itself, allocating the action range across the pump branch, the idle gap
and the generation branch in proportion to their widths at the current head
\cite{krasowski2023provably}. The agent therefore cannot express an
undeliverable commitment, and the projection is left to catch only what the
mask cannot anticipate, chiefly the drift of head within the hour. The DA and RT layers each carry their own mask, one acting on the hourly commitment and the other on the five-minute dispatch.
Second, in the
configuration reported here it is skipped on the terminal-balancing hour
$h\!=\!23$, whose commitment is computed to land SoC exactly at target, so
projecting it would perturb that balance. The projection subsumes the penalty, which we nonetheless keep switched on so
that it still charges for whatever the projection leaves behind. In the
taxonomy of provably safe reinforcement learning \cite{krasowski2023provably}
this is an \emph{action projection}, and terminal restoration
(Sec.~\ref{sec:terminal_restore}) is an \emph{action replacement} driven by
a fail-safe controller; retaining a penalty alongside such a mechanism rather
than in place of it is what that study recommends. 

\vspace{-2mm}
\subsection{Terminal State-of-Charge Restoration}
\label{sec:terminal_restore}

The DA agent's terminal-hour balancing pins the \emph{projected}
(zero-deviation) end-of-day SoC to $S^{\text{tgt}}$, but the RT agent's
deviations move the \emph{actual} storage away from that plan. We close the
residual gap by enforcing the target directly rather than by a reward term,
which would over-constrain the executor and, because the DA reward carries the
realized RT settlement, teach the planner to \emph{under-commit}
generation. The RT action is
overridden by a deterministic closed-loop dispatch: at each remaining interval
we compute the flow required to close the residual gap, invert the plant
physics (Eq.~\ref{eq:hydropower}) to the corresponding MW setpoint, and project
it onto the operating region. The override bypasses the $\pm\Delta_{\max}$
deviation cap, but the resulting deviation is settled at the realized RT-LMP,
so its cost or benefit is a \emph{true market settlement} rather than a tuned
penalty, and it reaches earlier actions through the standard Bellman backup.

Two details make the mechanism work. \rev{First, the controller drives
storage to a fixed target rather than to whatever level the day-ahead plan
projects, and that matters because the projection is unreliable: the hour-23
override lands it on target only when the required correction is itself
deliverable, leaving $\tilde S_T = 0.2491 \pm 0.0616$ with only $44$ of the
$144$ agent-days inside a percentage point and a worst day at $0.519$ where the
plan has filled the reservoir. Inheriting those projections would mean
inheriting their errors.} That is \rev{more than twenty times} looser than the realized error,
\new{so the day-ahead layer alone does not settle the end-of-day constraint,}
and an inherited target would pass the residual straight through. Second, engagement is adaptive.
\emph{Adaptive restoration} always runs over the final $N_0=12$ intervals and
engages earlier, up to $N_{\max}=72$ intervals before the end, whenever
closing the residual gap would require more than a fraction $\kappa$ of the
maximum feasible per-interval storage change at the current head. A fixed
window would instead have to be wide enough for the worst day, seizing
real-time authority on every day to protect against a rare one.

The margin trades lead time against real-time authority. We use
$\kappa=0.9$ throughout: a smaller margin would fire the trigger on nearly
every step and leave the controller in charge all day, while the remaining
tenth absorbs the contraction of per-interval capability as head moves.

Over the $144$ agent-days of Table~\ref{tab:paired_eval} the realized
end-of-day SoC is \rev{$0.2496 \pm 0.0023$}, \new{a \rev{twenty-sixfold} reduction in standard
deviation \rev{against the projected spread},} with \rev{$142$} landing within half a
percentage point of target and the three exceptions missing by $0.6$, $1.0$
and $2.2$ percentage points. Engaging at the last responsible moment spreads
the unwind over as many intervals as the gap requires, rather than a single
final hour.

\vspace{-2mm}
\subsection{Synthetic Price Model}
\label{sec:synthetic_lmp}
\vspace{-0mm}
We use synthetic, PJM-inspired LMPs for both training and evaluation.
The DA-LMP is piecewise constant per hour, with weekday/weekend and
peak/off-peak premia over a flat base, Student's-$t$ noise for fat tails, and
Bernoulli-triggered log-normal spikes and dips; the DA forecast adds Gaussian
noise. The RT-LMP is the DA-LMP plus a mean-reverting Ornstein-Uhlenbeck
process \new{at a fixed volatility
$\sigma_{\text{OU}}$.} \rev{The spikes and the real-time deviation stand in for the net-load ramps and forecast errors that wind and solar variability introduce, the dominant source of short-horizon price movement at high renewable penetration.}
The process is discretized at the 5-minute step and reset to zero at every
hour boundary. That reset is consequential rather than incidental: it makes
hours independent draws, so the real-time layer faces no intra-hour round trip
and no predictable inter-hour unwind, which is why the margin measured in
Sec.~\ref{sec:rt_value} is small. The RT-LMP forecast the agent observes
carries noise growing with horizon. Parameter values, the forecast noise
schedule and the statistics behind the reset claim are given in the
preprint \cite{park2026vspsharxiv}.

\vspace{-1mm}
\section{Numerical Results}
\label{sec:results}

\subsection{Experimental Setup}
\label{sec:results_setup}

The plant is the $100$\,MW unit of Sec.~\ref{sec:plant_model}, with
the operating-region curves of Sec.~\ref{sec:vsph_intro}. Episodes span 24
hours; the DA agent commits 24 hourly steps and then executes the frozen RT
agent on the cleared schedule. Each configuration is trained with three
independent DA seeds and scored on the same $48$ out-of-sample price days,
giving $144$ \emph{agent-days}. During training, periodic evaluation on a fixed
held-out seed band drives best-checkpoint selection.
That evaluation uses $120$ price days at each checkpoint, and
checkpoints are taken every $200$ day-ahead episodes and every $100$
real-time episodes, giving $10$ and $15$ respectively. The
budget matters. \new{Daily net revenue has a standard deviation of
$\$4{,}659$ across days within a single agent, so a checkpoint scored on $30$
days carries a standard error of $\$851$, larger than most of the effects the
ablations are meant to resolve, and taking the maximum over many such readings
selects the luckiest reading rather than the best policy. Widening the
evaluation to $120$ days halves that standard error, to $\$425$.} Unless stated otherwise, final numbers are
reported as the mean over an \emph{out-of-sample} set of 48 evaluation
price days, \new{drawn from a seed band disjoint from the selection band and
from a generator stream independent of the one producing training days,} and paired-system results are further averaged over
independent training seeds to report reproducibility. All numbers come
from evaluation, not the training reward. The seed bands are disjoint, so training, checkpoint selection, and out-of-sample evaluation never share a price day.

Three configuration choices matter for reproducibility rather than for performance. Best-checkpoint selection runs on a \emph{fixed} held-out seed band of deterministic price days, so the selection metric is comparable across checkpoints; selecting on randomly drawn days instead selects on evaluation noise. Rewards are normalized online and policy and critic gradients are clipped, without which DA training occasionally diverges at the reward scales that settlement cash flows produce \cite{haarnoja2018soft}. Finally, every reported number is a 48-day out-of-sample mean averaged over independent training seeds, and comparisons are paired on identical price days.

\vspace{-2mm}
\subsection{RT Controller: Schedule Diversity Drives Generalization}
\label{sec:results_rt_diversity}

The RT agent's training distribution is a design variable in its own right. An
agent's training-time evaluation curve is computed on \emph{its own} schedule
distribution, so it cannot compare agents trained on different mixes; we
therefore score every RT policy on a common per-class benchmark of $30$
episodes per class under the deployment configuration.

Concentrating round-2 training on the DA agent's own rollouts is intuitively
appealing, since the executor will only ever be handed such schedules, but it
does not pay. The proposed policy, trained on the seven synthetic classes
\emph{plus} a cache of $2{,}000$ DA rollouts, earns \$260/day more than an
otherwise identical policy trained \emph{exclusively} on that cache, and more
on every one of the seven classes in which the executor has discretion, which a
sign test resolves at $p\!=\!0.008$. The decisive comparison is the DA-rollout
class itself, where the specialized policy trails by \$261/day. The synthetic
classes therefore act as a regularizer that improves performance \emph{on the
target distribution} rather than as insurance bought at that distribution's
expense, a caution against narrowing an executor's training distribution onto
its deployment distribution once the latter is available.

\vspace{-2mm}
\subsection{From Penalty to Projection}
\label{sec:penalty_to_projection}

The rewrite-distance penalty $\lambda_{\text{op}} d_h^{\text{op}}$
(Eq.~\ref{eq:opclip_pen}) at $\lambda_{\text{op}}\!=\!\$50$/MW-substep does
push the median rewrite distance down, but it does not eliminate it reliably:
because a sub-minimum commitment is often profitable at settlement
(Sec.~\ref{sec:opclip_finding}), the soft penalty and the settlement
gradient pull against each other, and the surviving rewrite distance varies strongly with the training seed. On the 48-day out-of-sample set a
penalty-trained DA retains a median of \rev{$74$}\,MW-substeps under the present
recipe (Table~\ref{tab:paired_eval}, row \textbf{Y}), with some seeds clipping heavily and others not
at all. The operating-region \emph{projection} of
Sec.~\ref{sec:projection} removes this variance. Every commitment is snapped
onto the operating region before settlement, so the median rewrite distance
falls to $0$\,MW-substep. It does so while \emph{increasing}
revenue by \rev{\$3{,}313}/day, rather than trading revenue for
feasibility (Table~\ref{tab:ablations}). Feasibility
thus becomes \rev{a property the mechanism enforces} rather than an outcome the policy
may or may not learn.

\vspace{-2mm}
\subsection{Hierarchical System Performance}

Table~\ref{tab:paired_eval} reports the proposed paired system against a
penalty-based configuration and a zero-deviation ablation, all evaluated
on the same 48 out-of-sample price days. In that table DA, RT and Net are
settlement components in \$/day, SoC$_T$ is the mean actual end-of-day SoC
against a target of $0.25$ with the standard deviation pooled over all
$144$ agent-days, and op-rewr is the median operating-region
rewriting in MW-substeps/day. Net is market cash, $\mathrm{DA}+\mathrm{RT}$,
with no penalty wedge: the operating-region term is zero for every arm that
enforces feasibility structurally, so the reported revenue is settlement
the plant actually earned. Row \textbf{P} is averaged over three
independent training seeds, with the standard deviation on Net.
Every row is a complete pipeline retrained end to end with the named
change applied at each stage that it touches, rather than a single stage
swapped at the end; agents are training-process dependent, so reusing an
upstream agent from a different configuration would leave most of the
mechanism in place. The rows are:
\begin{description}
\setlength\itemsep{0pt}
\item[P] \emph{Proposed}: projection-trained DA + constructive-schedule, terminal-restoration RT. \rev{Mean over three independent seeds.}
\item[P\textsubscript{0}] \emph{Proposed DA + zero-deviation RT}: replaces the learned RT at evaluation time with an executor that follows the DA schedule and performs terminal restoration only.
\item[Y] \emph{Feasibility by penalty}: the operating-region penalty in place of both the action mask and the projection, retrained end to end. This is the configuration the battery literature would adopt, and it is the comparison the projection has to beat.
\new{\item[F] \emph{Perfect foresight}: the best deliverable plan for the same plant on the same days with the entire price path known in advance. Not a policy; a reference point. Its SoC spread is across days rather than seeds.}
\end{description}

\begin{table}[t]
\centering
\caption{Paired system average daily market cash.}
\label{tab:paired_eval}
\footnotesize
\setlength{\tabcolsep}{1pt}
\begin{tabular}{lrrr@{}lcr}
\toprule
& \textbf{DA} & \textbf{RT} & \multicolumn{2}{c}{\textbf{Net}} & \textbf{SoC$_T$} & \textbf{op-rewr} \\
\midrule
\textbf{P} (proposed) & \rev{$+$7{,}486} & \rev{$+$97} & \rev{\textbf{$+$7{,}583}} & \rev{{\tiny$\pm$367}} & \rev{$\mathbf{0.2496{\pm}.0023}$} & \textbf{0} \\
P\textsubscript{0} (P, zero-dev) & \rev{$+$7{,}486} & \rev{$-$74} & \rev{$+$7{,}412} & \rev{{\tiny$\pm$316}} & $0.2495{\pm}.0025$ & 0 \\
\textbf{Y} (penalty only) & \rev{$+$3{,}521} & \rev{$+$749} & \rev{$+$4{,}270} & \rev{{\tiny$\pm$1{,}146}} & $0.2500{\pm}.0006$ & \rev{74} \\
\midrule
\new{\textbf{F} (foresight)} & \new{$+$11{,}261} & \new{$+$1{,}073} & \new{$+$12{,}334} & & \new{$0.2488{\pm}.0024$} & \new{0} \\
\bottomrule
\end{tabular}
\end{table}

The proposed method earns more and is feasible. Against
configuration \textbf{Y} it gains \rev{$\$3{,}313$}/day. Because each arm retrains its own
agents, the unit of replication is the agent rather than the evaluation day.
The 48 days are not independent samples of a configuration change, since all
of them are produced by the same three agents. 
The mechanism is visible in the components.
Day-ahead settlement falls from \rev{$\$7{,}486$} to \rev{$\$3{,}521$} while
undeliverable commitment rises from $0$ to \rev{$74$} MW-substeps/day, and the
penalty arm's apparently healthy real-time settlement of \rev{$+\$749$} is not
skill but the executor buying back power the plant could not deliver.
\new{Row \textbf{F} asks what the same plant could have earned on these days
with the whole price path known in advance. We co-optimize the $24$ hourly
commitments and the $288$ dispatches by trust-region successive mixed-integer
programming, subject to the operating region, the deviation cap, the ramp
limit, the storage band and the same terminal target, and require every
commitment to be one the plant then follows. Each solution is replayed through
the simulator and settled on the dispatch actually executed, so it is an
achievable plan rather than an artifact of the linearization: rewriting and
undelivered commitment are both zero and terminal SoC lands at $0.2488$. The
proposed system captures \rev{$61.5\%$} of this benchmark, a median of \rev{$67.0\%$}
across days, and never exceeds it on any day. Successive programming converges
to a local optimum, so the true optimum is at least as large and the capture
ratio at most as high. More than four fifths of the shortfall is day-ahead
settlement: with foresight the benchmark commits far more aggressively than the
learned planner dares.}

Enforcing feasibility in the mechanism is therefore worth roughly \rev{$53\%$} of
day-ahead revenue in this setting, and the penalty formulation does not
approach it. Its spread across training seeds is also \rev{three times} the
proposed method's (\rev{$\$1{,}146$ against $\$367$}), which is the reproducibility
cost of leaving feasibility to the reward.
 The proposed
system also lands end-of-day SoC at \rev{$0.2496 \pm 0.0023$}, \new{within}
half a percentage point of target on \rev{$142$} of $144$ agent-days,
and it emits a median of $0$ MW-substep/day of operating-region rewriting
versus \rev{$74$}.

The learned RT agent is a modest contributor beside a day-ahead layer that
supplies most of the revenue; replacing it with a zero-deviation executor on
the same plan (\textbf{P\textsubscript{0}}) costs \rev{$\$172$}/day. The RT
settlement column does not measure arbitrage skill on its own, since it also
collects involuntary operating-region rewrites and the restoration dispatch,
which is why \textbf{P\textsubscript{0}} records a nonzero value despite
submitting no discretionary bids. Sec.~\ref{sec:rt_value} isolates it.

\begin{figure}[!t]
\centering
\includegraphics[width=\linewidth]{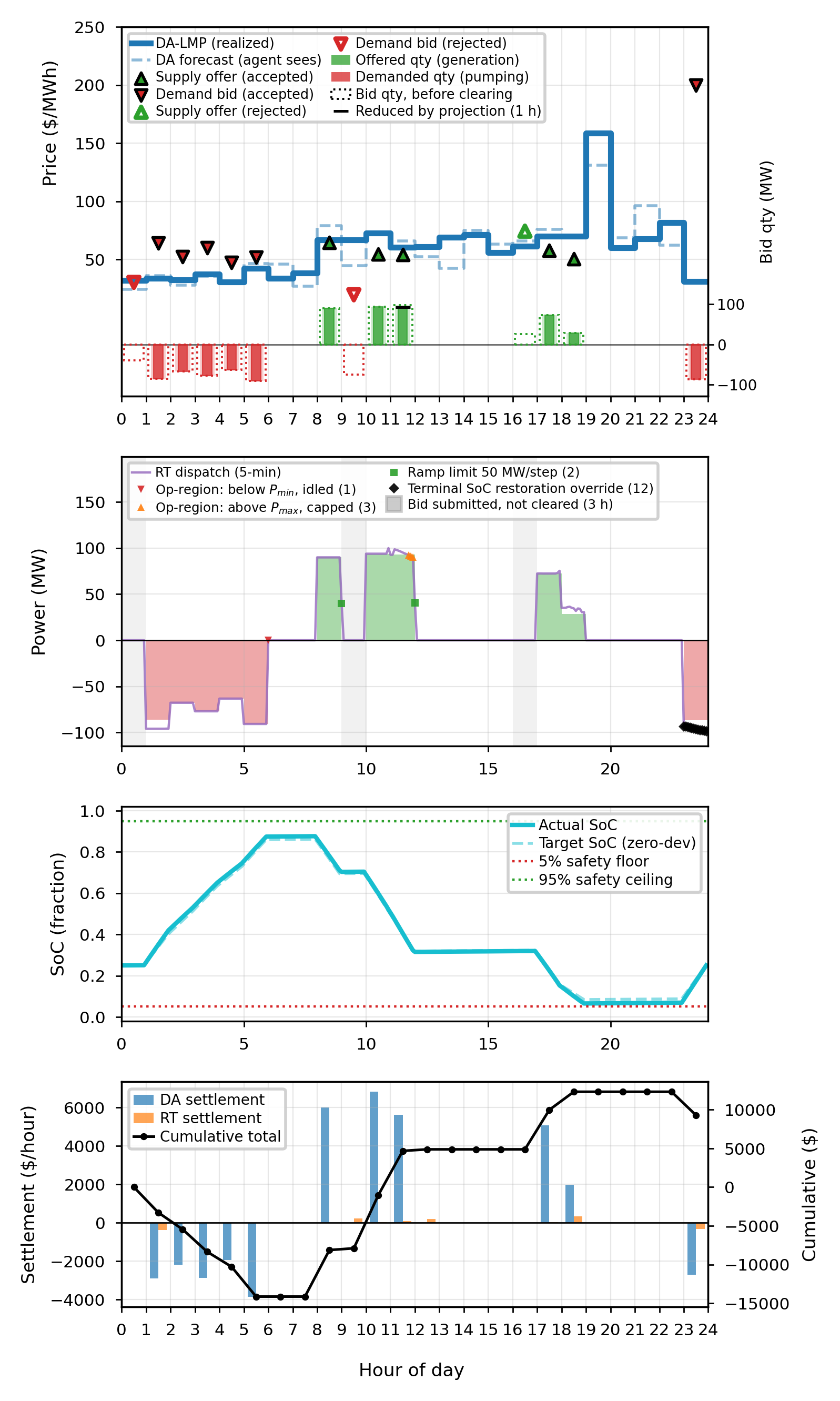}
\vspace{-7mm}
\caption{Panel 1: DA bids vs.\ realized DA-LMP; filled markers indicate
accepted bids. Panel 2: hourly DA commitments (bars) and 5-min RT
dispatch (line). Panel 3: SoC trajectory with the 5\%/95\% safety
rails. Panel 4: per-hour DA and RT settlement breakdown with
cumulative total.}
\label{fig:hier_day}
\end{figure}

Figure~\ref{fig:hier_day} shows the method on a representative
high-arbitrage day: bids clustered around the forecast, a charge--discharge
commitment profile, five-minute dispatch tracking it with small upward and
downward corrections against the real-time spread, and storage returned to
target by restoration.

\vspace{-1mm}
\subsection{Sensitivity to Each Design Choice}
\label{sec:ablations}

A combined result cannot attribute credit, so we ablate each mechanism
individually. Every arm branches from the \emph{same} upstream artifacts as the
proposed method, changes exactly one variable, trains three independent DA
seeds, and is scored on the same 48 out-of-sample days
(Table~\ref{tab:ablations}).

\begin{table}[t]
\centering
\caption{Single-feature ablations, columns as in
Table~\ref{tab:paired_eval} and $\Delta$Net relative to row \textbf{P}.}
\label{tab:ablations}
\footnotesize
\setlength{\tabcolsep}{3pt}
\begin{tabular}{llrcr}
\toprule
\textbf{Variable} & \textbf{Setting} & \textbf{$\Delta$Net (\$/day)} & \textbf{SoC$_T$} & \textbf{op-rewr} \\
\midrule
\multicolumn{5}{l}{\emph{Proposed method (P)}} \\
\quad --- & --- & --- & $0.250$ & 0 \\[2pt]
\multicolumn{5}{l}{\emph{DA feasibility}} \\
\quad Op-region & penalty only & \rev{$\mathbf{-3{,}313}$} & $0.250$ & \rev{\textbf{74}} \\
\quad DA action mask & removed & \rev{$-747$} & $0.250$ & 0 \\[2pt]
\multicolumn{5}{l}{\emph{RT layer}} \\
\quad RT action mask & removed & \rev{$-697$} & $0.250$ & 0 \\
\quad RT policy & zero-deviation & \rev{$\mathbf{-172}$} & $0.250$ & 0 \\[2pt]
\multicolumn{5}{l}{\emph{Pipeline}} \\
\quad Alternation & one round only & \rev{$-1{,}160$} & $0.250$ & 0 \\
\bottomrule
\end{tabular}
\end{table}

The feasibility mechanism dominates. Replacing the projection and the
action mask by a penalty costs an order of magnitude more than any other
change and is negative on every training seed. \rev{Removing the day-ahead
mask costs $\$747$ and is negative on every seed as well. The remaining two
arms, the real-time mask and the alternation round, average $-\$697$ and
$-\$1{,}160$ but are not consistently signed across seeds, so we report the
averages without reading a direction into them.}

A change made to the dispatch layer is partly hidden in system revenue,
because each arm's bidding layer is retrained against its own dispatch layer
and absorbs the difference by committing a schedule that needs less
correction. The per-class benchmark of Sec.~\ref{sec:results_rt_diversity} instead
scores every dispatch policy against the same fixed schedule distribution, and
there the day-ahead-specialized policy trails the proposed mix on every class
in which the executor has discretion. We therefore score dispatch-side changes
on the benchmark and reserve system revenue for bidding-side and
pipeline-level changes.

\vspace{-2mm}
\subsection{Marginal Value of the Real-Time Layer}
\label{sec:rt_value}

Is a learned dispatch layer needed at all, given a well-trained bidding
layer and the restoration controller of Sec.~\ref{sec:terminal_restore}? We
run, on identical price days and against the same day-ahead plan, the trained
dispatch policy and a zero-deviation policy, so the comparison is
storage-neutral. \new{Both arms share the same day-ahead agents, so the day is a valid unit of replication
and we pair over the 48 days.}

Day-paired over $48$ days and three DA seeds, the learned executor is
worth \rev{$+\$167$/day ($t\!=\!4.52$), ahead on $40$ of $48$ days and
$2.2\%$} of market cash. The gain is real trading rather than restoration
accounting. It earns \rev{$+\$389$/day in discretionary deviations and pays
$\$296$/day to unwind them, against $-\$156$ and $+\$82$ for the
zero-deviation executor, whose settlement is involuntary rewriting and
restoration rather than discretionary trading}
(Table~\ref{tab:paired_eval}).

\vspace{-1mm}
\section{Conclusions}
\label{sec:conclusions}
\vspace{-1mm}
\rev{Variable-speed pumped storage modulates power continuously in both
directions, a capability of growing value as wind and solar displace
dispatchable generation,} but it can do so only inside a head-dependent
nonconvex operating region, and it must bid into two settlements whose
decisions define each other. We cast this as a hierarchical RL problem and break the circular dependence between the layers by
training the dispatch layer against a constructive generator of deliverable
schedules before any bidding policy exists. Feasibility is then enforced in the mechanism rather than in the reward: action masking and projection hold every commitment inside the operating region, and an adaptive fail-safe controller returns storage to its contracted end-of-day level at market prices.

\vspace{-1mm}
\bibliographystyle{IEEEtran}
\bibliography{refs}

\end{document}